# Probabilistic graphical model based approach for water mapping using GaoFen-2 (GF-2) high resolution imagery and Landsat 8 time series

**Luyan Ji** [1], **Jie Wang** [2], **Xiurui Geng** [3] **and Peng Gong** [1,4,5]*

[1] Ministry of Education Key Laboratory for Earth System Modelling, Department of Earth System Science, Tsinghua University, Beijing 100084, China; E-Mail: jily@mail.ustc.edu.cn

[2] State Key Laboratory of Remote Sensing Science, Institute of Remote Sensing and Digital Earth, Chinese Academy of Sciences, Beijing, 100101, China; wangjie@radi.ac.cn

[3] Key Laboratory of Technology in Geo-Spatial information Processing and Application System, Institute of Electronics, Chinese Academy of Sciences, Beijing 100190, China; E-Mails: gengxr@sina.com

[4] Department of Environmental Science, Policy and Management, University of California, Berkeley, CA, 94720, USA

[5] Joint Center for Global Change Studies, Beijing 100875, China

* Correspondence: penggong@tsinghua.edu.cn; Tel.: +86-010-627-72750

**Abstract:** The objective of this paper is to evaluate the potential of Gaofen-2 (GF-2) high resolution multispectral sensor (MS) and panchromatic (PAN) imagery on water mapping. Difficulties of water mapping on high resolution data includes: 1) misclassification between water and shadows or other low-reflectance ground objects, which is mostly caused by the spectral similarity within the given band range; 2) small water bodies with size smaller than the spatial resolution of MS image. To solve the confusion between water and low-reflectance objects, the Landsat 8 time series with two shortwave infrared (SWIR) bands is added because water has extremely strong absorption in SWIR. In order to integrate the three multi-sensor, multi-resolution data sets, the probabilistic graphical model (PGM) is utilized here with conditional probability distribution defined mainly based on the size of each object. For comparison, results from the SVM classifier on the PCA fused and MS data, thresholding method on the PAN image, and water index method on the Landsat data are computed. The confusion matrices are calculated for all the methods. The results demonstrate that the PGM method can achieve the best performance with the highest overall accuracy. Moreover, small rivers can also be extracted by adding weight on the PAN result in PGM. Finally, the post-classification procedure is applied on the PGM result to further exclude misclassification in shadow and water-land boundary regions. Accordingly, the producer's, user's and overall accuracy are all increased, indicating the effectiveness of our method.

## 1. Introduction

Water shortage and pollution problems are the sustained focus by people and government of the arid and semi-arid regions, such as the Beijing-Tianjin-Hebei Region of China. Intensive monitoring over water area will be of great help for relieving the water stress in these areas. The recent advancement of remote sensing technology makes it possible by providing satellite data with finer spatial resolution and quality [1]. It is particularly effective when using very high resolution satellite imagery to monitoring small rivers, which are more dispersedly distributed and seasonally varying.

Currently, the most common situation for high resolution satellite data is composed by a pair of images where the first one acquired by a multispectral (MS) sensor in the visible and near-infrared (VNIR) range, and the second acquired by a panchromatic (PAN) sensor in the visible range with a higher spatial resolution. High spectral resolution helps distinguish water from other land cover types, while high spatial resolution assists in identifying water through texture and shape information. Date fusion is an effective technique to integrate MS and PAN image for water mapping. Image fusion methods, which have been extensively studied these years, can be categorized as being performed at three levels, namely, pixel, feature and decision levels [2]. The first two level methods aim to produce fusion result as input to a classifier, while the decision-level fusion was performed to combine initial decisions to acquire improvements.

Due to the ease of implementation, pixel-level fusion is widely used in many applications [3-7]. The geometrical registration is of vital importance [2]. Mismatch causes artificial colors or features, and brings the skewbald effect in boundary and shade regions. Moreover, it is hard to balance the amount of spectral information preservation with the amount of high spatial resolution retention [6]. Recent development shows that decision-level fusion method is more flexible and less sensitive to misregistration [8,9]. It permits the integration of expert knowledge and other information, such as spectral features, spatial texture and other ancillary information, in the classification and post-classification processes [8]. Its usefulness in land-cover mapping and other application has been proven by many researchers [10-12].

In remote sensing water mapping, a number of techniques using optical imagery have been developed, which can be categorized into four basic types: (1)statistical pattern recognition techniques (including supervised [13-16] and unsupervised classification methods [17]), (2) linear unmixing [18], (3) single-band thresholding [19,20] and (4) spectral indices [21-25]. A big challenge for mapping water is how to differentiate water from shadows (including the cloud, mountain, building and tree shadow) and other low-reflectance ground objects. It is mainly due to their spectral similarity, particularly in the VNIR range (, i.e. all have very low reflectance in most cases). This is a common problem for water mapping by all spatial resolution image, and becomes more severe in high resolution image. One possible way to recognize shadow is by utilize the geometric relationship among the source, object, sensor and shadow [26]. To differentiate water from low-reflectance objects, data from different sources with a wider band range can be useful, such as the Landsat data. It is because water has a significantly strong absorption in the shortwave infrared (SWIR) band range while most low-reflectance objects have a relatively weaker absorption. In addition, how to extract small water with a size comparable to or less than the MS spatial resolution is another challenging problem.

Therefore, the aim of this study is to develop a knowledge-based method for water mapping on high resolution imagery. Instead of directly combining the PAN and MS original data as an input into classifier, we introduce the probabilistic graphical model (PGM) by fusing the initial water probability result from PAN, MS. In addition, the Landsat 8 time series is also added for further improve the classification accuracy. Finally, a post-classification procedure including shadow identification and local unmixing is performed.

**2. Study area and Data**

*2.1. Study Area*

Our study area is located in rural-urban fringe of Beijing, the capital of China in the North China Plain (figure 1). The average precipitation per year is approximately 510 mm, and the average temperature is around 11 °C. Autumn in Beijing is typically hot and dry. Many built-up regions exist in this area, includes residential, commercial and industrial areas. The non-built-up regions mainly include the cropland, forest, grassland, shrubland and water. Large water bodies, such as Huairou

Reservoir, Shunyi Olympic Lake and Chaobai River, are shown in figure 1. In addition, small rivers, canals, lakes, ponds are distributed throughout this region.

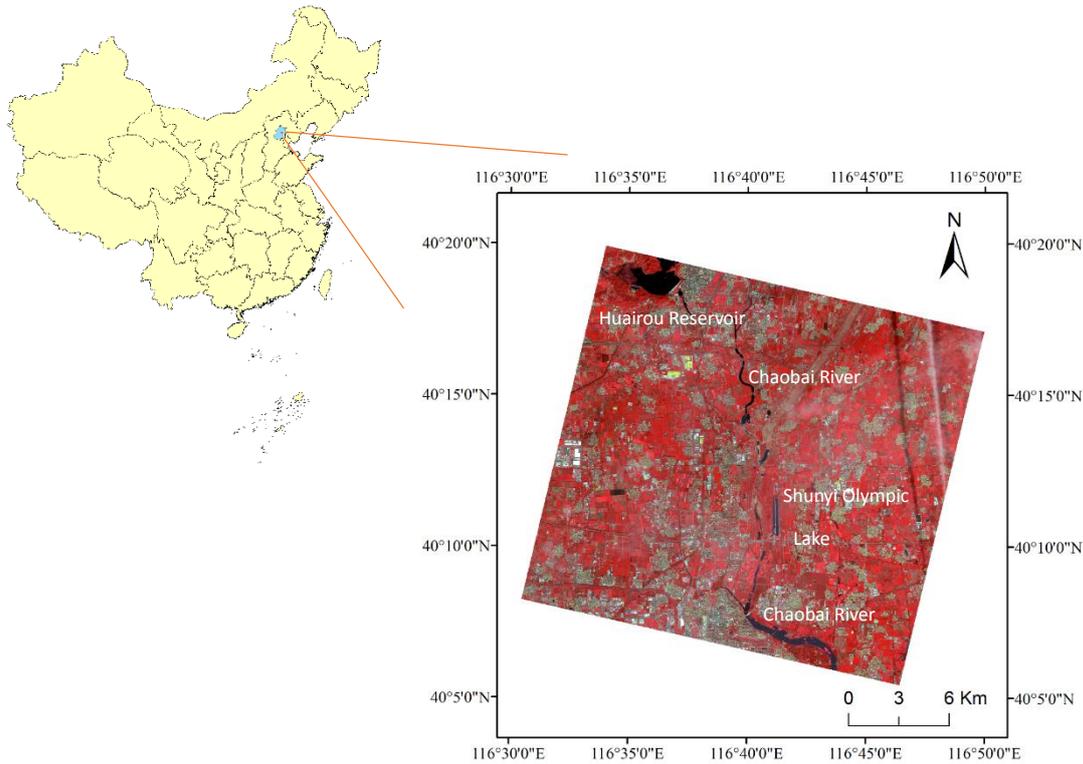

**Figure 1.** the study area used in this study. The Beijing City is shown in blue, and the image area is shown in red. The imagery is a false-color composite of NIR, red and green bands (Band 4, 3, and 2), which is acquired on 2 September 2015.

*2.2. Data*

2.2.1. GF data

In order to extract small water bodies in this area, the GF-2 high-resolution imagery is selected for water mapping. The GF-2 satellite was launched in 2008 and since then can provided sustainable high resolution imagery with high quality. The GF-2 data contains two types of imagery: one is the MS image with four bands in the VNIR range with a spatial resolution of 3.2 m; and the other is PAN image in the visible range with a spatial resolution of 0.8 m. The sizes for MS and PAN images are 8658×8399 and 34632×33594 respectively. The images are acquired on 2 September 2015 and with little cloud (figure 1). Both MS and PAN images are orthorectified using the rational polynomial coefficient (RPC) transformation with near neighborhood resampling method. The PAN image is registered to the control image with 100 ground control points (GCPs) and wrapped using the 1st polynomial with nearest neighbor interpolating method. Next, the MS image is registered to the PAN image with 119 GCPs with precision better than 0.6 pixels.

2.2.2. Landsat time series

All Landsat 8 reflectance images of this area (path=123, row=032) in year 2015 are collected from the United States Geological Survey (USGS) portal (http://earthexplorer.usgs.gov/). The Landsat 8 data contains 7 bands from VNIR to SWIR range with a pixel size of 30 m. Images with cloud or snow/ice cover in the study area are removed, and thus a total of 7 scenes are selected for final use. The days of year (DOYs) for the 7 images are 106, 122, 138, 186, 234, 250 and 266. The main reason we choose this data is because the Operational Land Imager (OLI) sensor carried by Landsat 8 includes SWIR bands. Since the liquid water has much stronger absorption in SWIR range than in VNIR range, water reflectance in SWIR is much lower than that in VNIR. Therefore, the two SWIR bands (1.560 - 1.660 μm, 2.100 - 2.300 μm) of Landsat 8 image will be of great help in differentiate water from other low-reflectance ground objects.

## 3. Method

*3.1. Difficulties of water mapping using GF high resolution imagery*

3.1.1. Confusion between shadows and water

For water mapping with optical remote sensing image, the shadows are always easily to be misclassified as water. It can be attributed to the fact that they both have relatively lower reflectance compared to the other ground objects, such as vegetation and bareland. The types of shadow in remote sensing imagery usually include cloud, mountain, building and tree shadows. In the high resolution image of our study area, there mainly exist two types of shadow, the building and tree shadow. Examples of building and tree shadows are plotted in figure 2, from which one can see that both shadows appear very dark in the MS and PAN image.

Further, we manually selected 100 pixels on the MS image for the building shadow, tree shadow and water respectively. The spectral signatures are plotted in Figure 3(a) to demonstrate the spectral overlaps between water and shadow. One can find that both water and shadow have very low reflectance in all four VNIR bands with strong spectral overlaps. This is why shadows are easily misclassified as water.

It should be noted here that a long white line by airplane can be visually observed in the upper-right part of the image (Figure 1). Since it is relatively thin and few water exists in this "cloud" and "cloud shadow" area, we have not done special processing on this area.

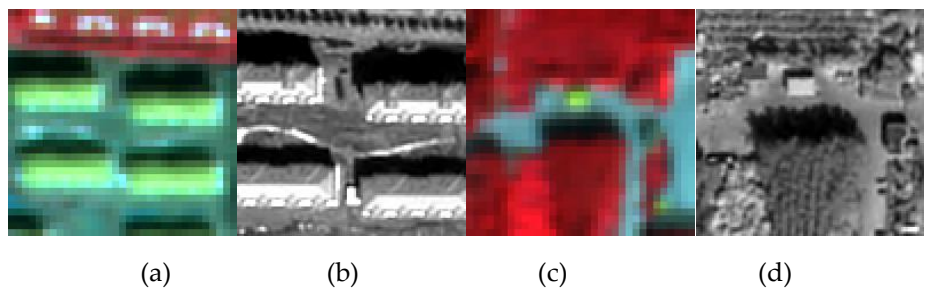

(a)          (b)          (c)          (d)

**Figure 2.** Illustration of the building and tree shadows in the MS and PAN image. (a) and (c) are the false color image of the MS image (R: $b_4$, G: $b_3$, B: $b_2$); and (b) and (d) are the grey image of the PAN image.

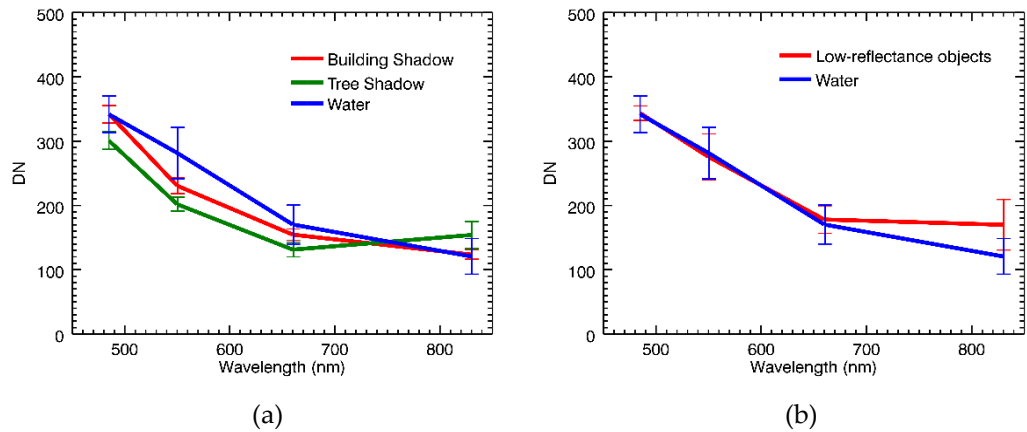

(a)                          (b)

**Figure 3**. (a) Comparison of spectral signatures (mean ± standard deviation) among building shadow, cloud shadow and water. (b) Comparison of spectral signatures (mean ± standard deviation) between low-reflectance ground objects and water. 100 signatures of each type were manually selected from the MS image.

3.1.2. Confusion between low-reflectance objects and water

Due to the limited band range, some ground objects with low-reflectance in the VNIR bands have a high probability to be detected as water by many classifiers. Similarly, we selected 100 pixels for the low-reflectance objects on the MS image and the spectral signatures are plotted in Figure 3 (b). Again, their spectra are seriously overlapping with water's, especially in the visible bands. An example of low-reflectance area is demonstrated in Figure 4. From the MS false color image, it is very difficult to determine whether this area is covered by water, even to an expert. Yet, according to the field survey (Figure 4(b)), this area is a field of farmland, which grows the American ginseng (*Panax quinquefolius*) and is covered with the black film all year around.

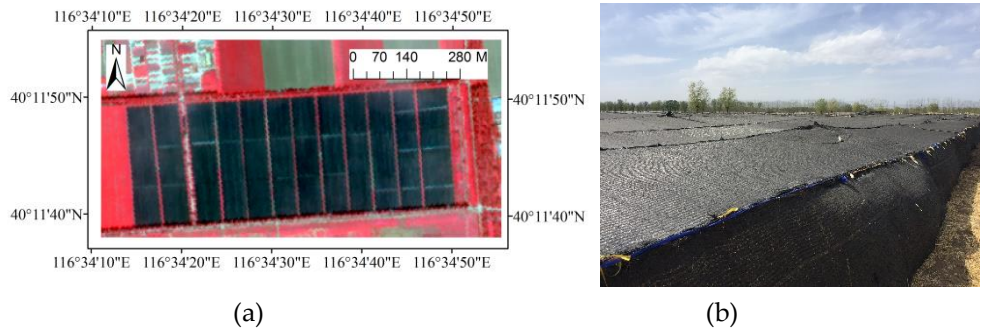

(a)                          (b)

**Figure 4**. An example of low-reflectance object in the MS image. (a) the false color image (R: $b_4$, G: $b_3$, B: $b_2$); (b) the field collected photo in this area.

(3) Extraction of the small water body

3.1.3. Extraction of the small water body

Detection of small water bodies with a size comparable to or less than the resolution of MS image is a very challenging problem. For example, the small river shown in Figure 5 is about 2.4 m in width, which is visible in the PAN image (3 pixels in width). However, it is hard to be discovered in the MS image due to the spectral mixing effect, so that would be easily misjudged into other land cover types (such as vegetation). Therefore, when using relatively easy combination rules between MS and PAN classification results, we are very likely to miss this kind of small rivers.

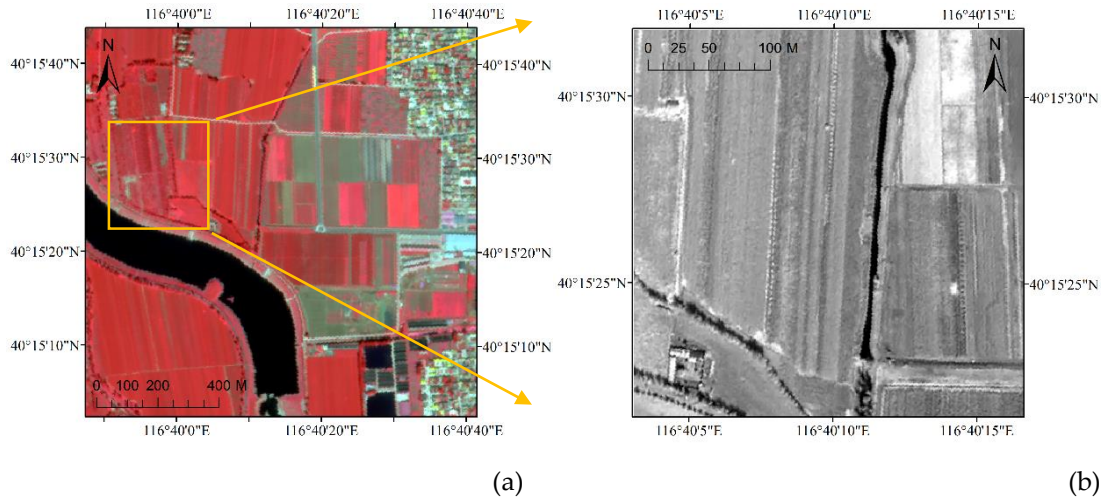

(a)                          (b)

**Figure 5.** Difference of the small river in the MS and PAN image. (a) the false color image of MS data (R: $b_4$, G: $b_3$, B: $b_2$); and (b) the corresponding PAN image.

*3.2. Pixel-level image fusion method (the PCA based method)*

At pixel level, the PCA based method is one of the most widely used methods in image fusion methods[7]. The procedure of PCA image fusion method is as follows: 1) Performing a PCA transformation on the MS data; 2) Replacing the first principle component (PC) with the PAN image, and scaling the PAN to match the first PC, so no distortion of the spectral information occurs. It assumes that the first PC band is a good estimate of the PAN; 3) Performing an inverse transform; and 4) Resampling the multispectral data to the high resolution pixel size. In this paper the nearest neighbor resampling method is applied.

It should be noted here that, before performing the PCA fusion, the MS should be geometrically registered to the PAN image. Like other pixel level based methods, the PCA fusion method has an extremely high requirement on the matching accuracy. When the matching accuracy is lower, the skewbald in object boundary and shade region is very like to appear.

*3.3. Supervised classification for water mapping on the MS and PCA fusion image*

Currently, the support vector machine (SVM), which can provides good classification results from complex and noisy data, is one of the most frequently used supervised classification method for land cover mapping [27,28]. For both MS and PCA fusion image, SVM using a radial basis function kernel is used for water mapping. The training samples of four land cover types, the vegetation, soil, impervious surface and water, are manually selected from the PCA fusion image first. For fair comparison, the same set of training samples is transferred to the MS image for classification. Moreover, we use the same parameter setting of SVM for both data sets. In addition, SVM performs probability estimation by solving a restricted minimization problem. Therefore, it can produce probability information for each land cover type, which will be utilized as the probability estimation in the latter process on PAN image.

*3.4. Water identification on Landsat image*

According to the absorption coefficient spectrum, liquid water has a very weak light absorption in the visible region, but an extremely strong absorption in the SWIR region. Because of this fact, surface water in the Landsat image usually has a lower reflectance in the SWIR bands than in the visible bands. Therefore, water pixels can be differentiated from other ground objects by simply comparing the value of visible bands and SWIR bands as the water index used in [26],

$$WI_i = \begin{cases} 0, & \text{if max\{visible bands\}} \leq \text{max \{SWIR bands\}} \\ 1, & \text{if max\{visible bands\}} > \text{max \{SWIR bands\}} \end{cases}. \qquad (1)$$

where $i = 1, 2,…,M$ ($M$ is the number of Landsat images, and $M=7$ in this study). Compared to the water index in [26], the NIR band is abandoned here because the reflectance difference between SWIR bands and visible bands is larger than that between NIR bands and visible bands. Next, the probability of water for Landsat image is estimated by

$$P(LAN = \text{water}) = \frac{1}{M}\sum_i WI_i. \qquad (2)$$

*3.5. Object-based method for water mapping on the PAN image*

The water body is usually spatially compact, and pixels within a water body are relatively homogeneous both spectrally and spatially. Therefore, we adopt the object-based method for water mapping on the PAN image. The segmentation on the PAN image can be directly performed by using the clustering method. However, due to the lack of rich spectral information, different land covers could have same values, while the same land covers may have different values. Therefore, the spatial information should be exploited to enhance segmentation result.

The morphological profiles (MPs) based on mathematical morphology are applied to extract the structural information [29,30]. To better fit several shapes in the image, three types of structuring elements (SEs) are used, the horizontal line, the vertical line and the square with a fixed size of 4×4 pixels[29]. Each shape of the SE is used twice, i.e. for one opening operation and one closing operation, for structural information extraction. In addition, the square shaped SE with an increasing size of 6×6 and 8×8 are also computed. As a result, a total of 10 profiles are constructed through the repeated use of SEs with different shapes and sizes. Next, the segmentation is simply carried out by K-Means on the PAN and MPs data with the number of clusters setting to 8. Then the probability of a segment covered by water can be estimated by

$$P(PAN = \text{water}) = \frac{N(\text{DN} < T_{PAN})}{N}, \qquad (3)$$

where $N$ is the total number of pixels in the segment, and $N(\text{DN} < T_{PAN})$ represents the number of pixels with digital value (DN) smaller than a given threshold $T_{PAN}$.

*3.6. Probability estimation for the shadow object*

Generally, an object with certain height can produce a shadow adjacent to it. Supposing the height of a pixel ($X_O$, $Y_O$) is $h$, the position of its shadow pixel ($X_S$, $Y_S$) in the image can be expressed as[31]

$$X_S = X_O + a \cdot \frac{h}{r}, \qquad (4)$$

$$Y_S = Y_O + b \cdot \frac{h}{r}. \tag{5}$$

where r is the spatial resolution of the image; a and b are two coefficients only related to the sun elevation and azimuth angles, and viewing elevation and azimuth angles. For our GF-2 PAN image, *a*, *b* and *r* are all known. Therefore, given $X_O$, $Y_O$ and *h*, we can accurately calculate the shadow position.

Mostly, ground objects that have visible shadows in an image with 0.8 m resolution are building and tree. Thus, in order to produce the shadow area with (4) and (5), we need to extract the building and tree pixels in the PAN image first. Here, the building pixels are approximated by the impervious surface, and the tree needs to be distinguished from grass in the vegetation class. Yet, it is impossible to classify impervious surface and tree (or even vegetation) from the single PAN image or the MPs, so we employ the majority voting method for PAN segment classification by fusing the MS classification results. Since grass is more homogeneous than tree, the grass spectrum of MPs data is much flatter than tree's. We compute the standard deviation (STD) of the MPs data, and the PAN vegetation segment with an average STD value larger than a given threshold is classified as tree.

Though we can derive the positions of buildings and trees, their accurate heights are difficult to obtain from the PAN image directly. To set a better range for the building height, the ratio of impervious area is used to map out the high-intensive building area and low-intensive building area with a threshold of 30% (in a 101×101 pixel size window). The height ranges for the high-intensive and low-intensive building area are set to [3, 300m] and [3, 50m], respectively. For the tree height, the height range is uniformly set to [3, 50m]. Then, the potential shadow mask can be produced according to (4) and (5). Next, the proportion of a segment with potential shadow pixels can be computed by

$$p_{\text{shadow}} = \frac{N(\text{shadow mask}=1)}{N}, \tag{6}$$

where *N* is the total number of pixels in the segment, and $N(\text{shadow mask}=1)$ is the number of pixels being classified as shadow in this segment.

*3.7. Probabilistic graphical model for water mapping*

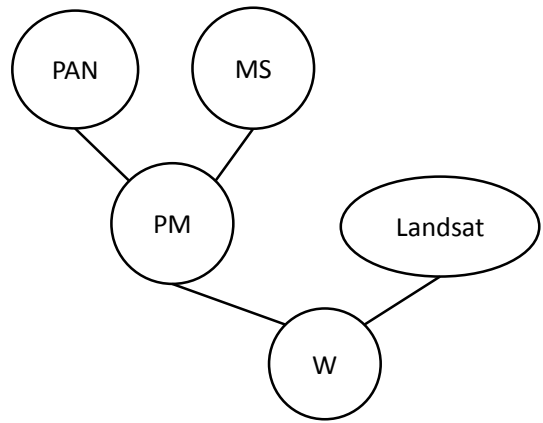

**Figure 6.** Illustration of the probabilistic graphical model for water mapping.

The probabilistic framework has a central role in scientific data analysis, machine learning and artificial intelligence [32,33]. In this study, the PGM for water mapping by combing the PAN, MS and

Landsat results is shown in Figure 6, where PM is the combined result of PAN and MS, and W is the estimation for real land cover (water or non-water). Accordingly, the random variables for the nodes are defined as *PAN*, *MS*, *LAN*, *PM*, and *W*. Then the marginal probability distribution of PM can be computed by

$$P(PM) = \sum_{PAN,MS} P(PM, PAN, MS) = \sum_{PAN,MS} P(PM | PAN, MS) P(PAN) P(MS) \quad (7)$$

where $P(PM, PAN, MS)$ is the joint probability distribution, and $P(PM | PAN, MS)$ is the conditional probability distribution (CPD), which is defined by

$$P(PM | PAN, MS) = \begin{cases} 1, & \text{if } PM = PAN = MS \\ 1 - S\left(\left[w/(n_1 r_{MS}) + p_{shadow}\right]/2\right), & \text{if } PM = PAN \neq MS \\ S\left(\left[w/(n_1 r_{MS}) + p_{shadow}\right]/2\right), & \text{if } PM = MS \neq PAN \\ 0, & \text{others} \end{cases} \quad (8)$$

where w is the size of the PAN segment, $r_{MS}$ (= 3.2) is the spatial resolution of the MS image, and n1 is a positive integer. The multiplication of $n_1 r_{MS}$ indicates the minimum size of a water segment that can be detected in the MS image. $S(t) = 1/(1+e^{-t})$ is the sigmoid function. The CPD definition in (8) indicates that if a segment has a larger size and contains more potential shadow pixels, we put more weight on the MS result, and vice versa.

Next, the marginal probability distribution $P(W)$ can be computed similarly by

$$P(W) = \sum_{PM,LAN} P(W, PM, LAN) = \sum_{PM,LAN} P(W | PM, LAN) P(PM) P(LAN), \quad (9)$$

where the CPD $P(W | PM, LAN)$ is defined as

$$P(W | PM, LAN) = \begin{cases} 1, & \text{if } W = PM = LAN \\ 1 - S(w/(n_2 r_L)) \cdot \delta(w - n_2 r_L), & \text{if } W = PM \neq LAN \\ S(w/(n_2 r_L)) \cdot \delta(w - n_2 r_L), & \text{if } W = LAN \neq PM \\ 0, & \text{others} \end{cases} \quad (10)$$

where $r_L$ (= 30) is the spatial resolution of the Landsat image, $n_2$ is an positive integer, and their multiplication, $n_2 r_L$ indicates the minimum size of a water segment that can be detected in the Landsat image. $\delta(t)$ is a 0-1 function defined as $\delta(t) = 1$ if $t \geq 0$, and $\delta(t) = 0$ if $t<0$. The CPD definition in (9) manifests that if a segment has a size larger than $n_2 r_L$, we should put more weight on the Landsat result. And the larger the size is, the more weight we should put on the Landsat result.

*3.8. Post-classification procedure for shadow removal*

As shown in figure 2, building and tree shadows exist on both MS and PAN images. As a result, misclassification between shadow and water still exist in the water mapping result by PGM. On the other hand, though the PAN image has a very high spatial resolution, spectral mixing problem still exist in the water-land boundary. With respect to these two problems, the post-classification method proposed in the reference [26] is utilized here. For a given water segment from PGM result, if the proportion of potential shadow pixels, $p_{shadow}$ is larger than 0.85, this segment is relabeled as a shadow. Finally, the local unmixing is performed at the water-land boundary.

**3.9 Validation**

The accuracy assessment is done by a comparison of validation samples, which are selected using the stratified sampling strategy as a guideline. For the classification result of PCA fusion image, 100 pixels are randomly picked from the vegetation, soil and impervious surface layer respectively, and 300 pixels are randomly selected from the water layer, such that a total of 600 sample pixels are selected. In order to maintain a fair comparison, the same set of samples is used to evaluate the classification result on the Landsat, MS and PAN image.

The land class for each validation pixel was identified manually so that it can be used as the reference data in the accuracy assessment. The identification is aided by using the multi-temporal high-resolution image from the Google Earth.

The main purpose of this study is water mapping, so all the vegetation, soil and impervious surface classes are integrated into a new class, non-water. And for each classification result, a confusion matrix of water and non-water classes was produced by comparing the classification results for the validation samples with the reference data. The producer's accuracy (PA), user's accuracy (UA), and overall accuracy (OA) are calculated for each confusion matrix to evaluate the agreement between classification results and the reference data.

**4. Results**

*4.1. Water mapping results using the PCA fusion, PAN, MS and Landsat image*

Table 1 tabulates the confusion matrix for the SVM classification result on the PCA fusion method. From Table 1, we can see that the SVM+PCA method has a very high PA (PA=99.1%) but a low UA (UA=76.9%). As a result, the overall accuracy is low (OA=88.2%). It indicates that the SVM+PCA method has over-classify the water class. Shadows and low-reflectance objects are misclassified as water by the SVM+PCA method.

Table 1, Confusion matrix for the pixel-based water and non-water classification result using the PCA fusion image.

|  |  | Reference |  |  |
|---|---|---|---|---|
|  |  | **Non-water** | **water** | **UA (%)** |
| **Classification** | **Non-water** | 299 | 2 |  |
|  | **water** | 69 | 230 | 76.9 |
|  | **PA (%)** |  | 99.1 | 88.2 |

The same happens to the object-based method using the PAN image based on (3), as shown in Table 2. Since the PAN image has a higher spatial resolution, the shadow or low-reflectance pixels are less mixed with the other land cover pixels. Thereby, the shadow or low-reflectance pixels are more likely to be misclassified as water. Accordingly, the UA and OA of the object-based method on PAN image are lower than those of the SVM+PCA method.

Table 2. Confusion matrix for the object-based water and non-water classification result using the PAN image.

|  |  | Reference |  |  |
| --- | --- | --- | --- | --- |
|  |  | Non-water | water | UA (%) |
| **Classification** | **Non-water** | 249 | 0 |  |
|  | **water** | 119 | 232 | 66.1 |
|  | **PA (%)** |  | 100 | 80.2 |

For fair comparison, we first resample the water probability result of SVM+MS result and Landsat to the spatial resolution of PAN. Next, the average water probability of each segment is computed. The confusion matrices for the object-based result using the MS and Landsat data are tabulated in Table 3 and 4 respectively. Compared with the SVM+PCA and PAN result, the UAs of both MS and Landsat images are increased a lot. It is mainly because MS and Landsat data contains more spectral bands, so some shadows, that has obvious non-water spectral characteristics (though with a low reflectance value) will not misclassified as water. However, the Pas of both data is decreased. For the MS data, the main reason is that some water pixels at the water-land boundary are misclassified as non-water class. And for the Landsat data, since it has a much coarse spatial resolution, water bodies with a small size (for example, less than 30m) are extremely difficult to detect. That's why PA of the Landsat image is the lowest. Yet, the overall accuracies of both data are all increased.

Table 3. Confusion matrix for the object-based water and non-water classification result using the MS image.

|  |  | Reference |  |  |
| --- | --- | --- | --- | --- |
|  |  | Non-water | water | UA (%) |
| **Classification** | **Non-water** | 339 | 12 |  |
|  | **water** | 29 | 220 | 88.4 |
|  | **PA (%)** |  | 95.0 | 93.2 |

Table 4. Confusion matrix for the object-based water and non-water classification result using the Landsat image.

|  |  | Reference |  |  |
| --- | --- | --- | --- | --- |
|  |  | Non-water | water | UA (%) |
| **Classification** | **Non-water** | 361 | 33 |  |
|  | **water** | 7 | 199 | 96.7 |
|  | **PA (%)** |  | 85.8 | 93.3 |

*4.2. Water mapping result based on probabilistic graphical model*

The accuracy assessment of our PGM is shown in Table 5. Compared with the PAN result in table 2, the UA has been significantly increased, while compared with the Landsat result in table 4, the PA has increased a lot. And compared with the MS result in table 3, both PA and UA have been slightly increased. In all, since PGM combines all the information from the PAN, MS and Landsat data, its overall accuracy is the highest (OA=94.2%).

Table 5 Confusion matrix for the classification result of the PGM.

|  |  | Reference | | |
| --- | --- | --- | --- | --- |
|  |  | Non-water | water | UA (%) |
| Classification | Non-water | 341 | 8 |  |
|  | water | 27 | 224 | 89.2 |
|  | PA (%) |  | 96.6 | 94.2 |

The increase in the PA of PGM to that of PAN and Landsat is mainly because large low-reflectance objects are rejected by PGM with additional Landsat information. For example, the farmland in Figure 4, is covered some kind of black film all year round, so its reflectance in VNIR is very low. As a result, the water probability estimated by both PAN and MS image is high, as shown in Figure 7 (b) and (c). However, since Landsat data has a wider band range, the water probability calculated by (2) on the Landsat time series is very low (Figure 7 (d)). By fusing the three results, the water probability estimated by PGM in this area is low (Figure 7 (e)).

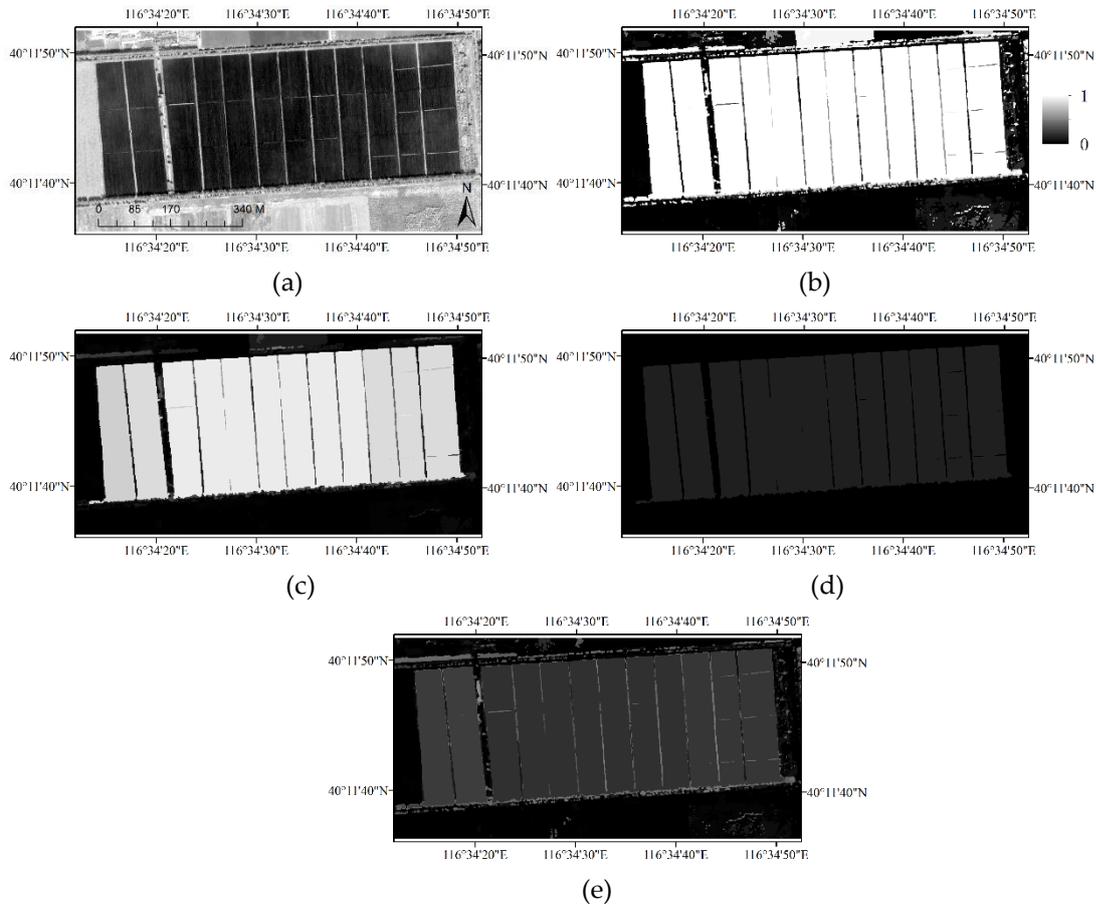

Figure 7, (a) the PAN image of a low-reflectance farmland planting the American ginseng. And the water probability of PAN (b), MS (c), Landsat (d) and PGM (e) result.

In the other aspect, by fully ulitizing the PAN information, small river with a size less than the MS spatial resolution can be detected by the PGM. For the small river introudced in Figure 5, the water probability of MS result is actually very low as shown in Figure 8(c). Yet, since the water probability of the PAN result is high (Figure 8(b)) and the river size is small, the final water probability esitmated by PGM is larger than 50% (Figure 8(e)). Therefore, the PGM has the ability for small river detection by adding more weight on the PAN result for small-sized water bodies.

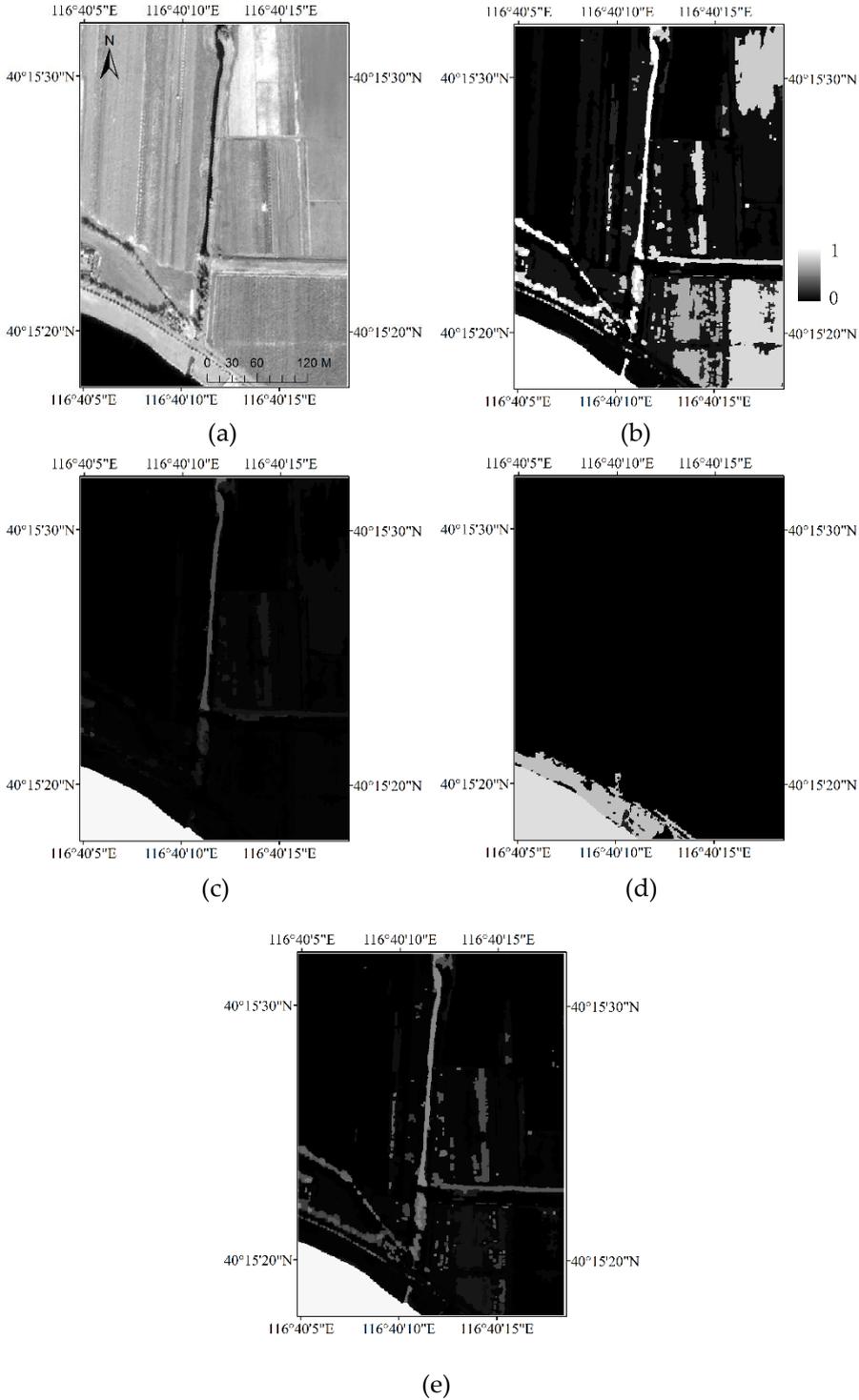

Figure 8, (a) the PAN image of a small river, and the water probability of PAN (b), MS (c), Landsat (d) and PGM (e) result.

*4.3. Results of the post-classification procedure*

From Table 5 we can see that the UA of the PGM method is relatively low compared with the high PA value. It is mainly because shadows universally exist in the PAN image, and are easily to be misclassified as water. Thereby, the shadow removal is a necessary procedure after water mapping. Figure 9 shows two example of shadow identification results by our post-classification method. Large building and tree shadows are all identified and excluded from the water class. The final confusion matrix is tabulated in Table 6. Compared with Table 5, the UA after post-classification process has been greatly improved, which is mainly attributed to the shadow removal. Moreover, the PA value is also slightly increased, which is contributed by the local unmixing in the post-classification process. Pixels in the water-land boundary with larger water fractions are relabeled as water again.

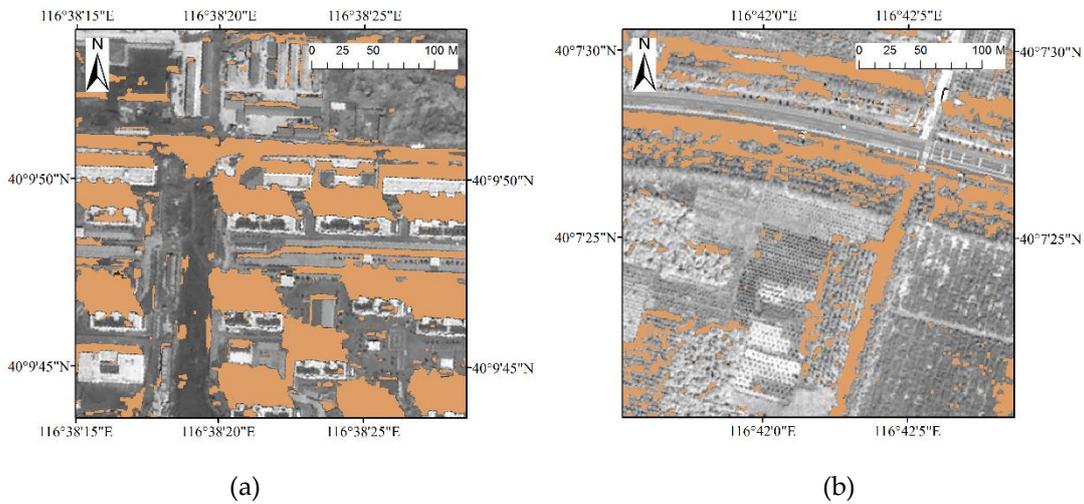

|     (a)     |     (b)     |

Figure 9, Results for building shadow (a) and tree shadow(b) identification.

Table 6 Confusion matrix for the classification result of the post-classification processing.

|  |  | Reference | | |
|---|---|---|---|---|
|  |  | **Non-water** | **water** | **UA (%)** |
| **Classification** | **Non-water** | 350 | 7 |  |
|  | **water** | 18 | 225 | 92.6 |
|  | **PA (%)** |  | 97.0 | 95.8 |

## 5. Discussion

Tables 1-5 reflect the advantage of PGM in the multi-sensor and multi-data integration. It can intelligently choose more reliable, accurate, and useful knowledge from multiple existing classification results for different objects. The CPDs in (7) and (9), which is key to PGM, are very similar to the commonly used weight coefficients. However, the weight values are usually fixed, but PGM can automatically adjust the CPD value for each individual object. In this paper, the adjust rule is based on the object size variable. Therefore, PGM can guarantee classification accuracies for water bodies with all kinds of size.

However, from Table 6 we can see that the UA of our method is relatively lower compared with PA, which indicates that a few non-water objects are still misclassified as water. There mainly exit two cases causing this kind of misclassification:

1) Small shadow adjacent with the true water body is more likely to be incorrectly labeled as water, since it will be identified with the same region index as the water body. One possible way to solve this problem is by splitting the current object into several smaller objects, and then utilizing the spectral differences to identify the shadow object. However, if the shadow object is too small to be recognized in the MS image, this method could still fail. Another possible way is to this problem is to accurately calculate the shadow area by using the accurate height information.

2) Small dark object could be mistakenly classified as water. Our method relies more on the MS/PAN result than the Landsat image if the object size is smaller. As a result, if a dark object with a size smaller than 30m and has no spectral difference to the water in VIS bands, it is easily categorized into water class mistakenly. More prior knowledge is required if we want to reject this kind of dark object.

Therefore, with more information from disparate sources, the classification accuracy can still be improved. Fortunately, PGM has no limitation on the number of data sources. If more data is added, we only have to increase the number of nodes and design reasonable CPDs accordingly. This shows the real power of PGM in the field of multi-data fusion.

## 6. Conclusion

In this research, we investigate the fusion of GF-2 MS, PAN image and the Landsat 8 time series for water mapping in an urban and rural fringe area. The main problems in this topic include: 1) how to take the full advantage of the MS, PAN and Landsat images for water extraction, including small water bodies; and 2) how to reduce the misclassification between water and shadow or other low-reflectance ground objects. To solve the problems, the object-based probabilistic graphical model (PGM) is used to fuse the initial PAN, MS and Landsat water estimations on a decision level. As a result, large sized dark objects which cannot be distinguished by PAN and MS data can be classified as non-water class. It is because they exhibit different spectral characteristics in the Landsat data and we have added more weight on the Landsat result by adjust the CPD in PGM. Moreover, small water bodies can also be identified by putting more weight on the PAN result. Moreover, a post-classification procedure is proposed by utilizing the geometric relationship between object and shadow for shadow removal. And we also relabel the land cover types for pixels in the water-land boundary by a local unmixing result. Experimental results show that our proposed method is effective in water mapping with high accuracy (PA=97% and UA=93%, and OA=96%). And it can directly be applied to synergistically combine multi-data from other sources.

**Acknowledgments:** This research is partially supported by the National Basic Research Program of China (973 Program) under Grant Number 2015CB953701.